\documentstyle[aps,prb,multicol]{revtex}

\newcommand{\beq}{\begin{eqnarray}}
\newcommand{\eeq}{\end{eqnarray}}
\newcommand{\n}{\nonumber}
\newcommand{\epi}{\epsilon_\infty}
\begin{document}
\draft
\title
{Johnson-Nyquist noise in films and narrow wires}
\author{ Misha Turlakov}
\address{ Department of Physics, University of Illinois, 1110 W. Green Street, Urbana, IL 61801}
\date{\today}
\maketitle

\begin{abstract}

The Johnson-Nyquist noise in narrow  wires having a transverse size smaller than
the screening length is shown to be white up to the frequency $D/L^2$ and to decay
at higher frequencies as $\omega^{-\frac{1}{2}}$. 
In two-dimensional films having a thickness smaller than the screening length, 
the Johnson-Nyquist noise
is predicted to be frequency independent up to the frequency $\sigma_{2D}/L$  
and to have a {\it universal} $1/\omega$ spectrum 
at higher frequencies.
These results are contrasted
with the conventional noise spectra in neutral and three-dimensional charged liquids.

\end{abstract}

\begin{multicols}{2}


 It is interesting to analyze the differences between charged and neutral systems
due to the long-range nature of the Coulomb interaction.
The role of the Coulomb interaction 
 depends crucially on the effective dimensionality
of the charged system. For instance, 
 in three-dimensional systems Coulomb interaction transforms 
the gapless density excitations of neutral liquids
(acoustic phonons) to gapped plasmons. Nevertheless, in one- and two-dimensional
systems plasmons remain gapless.   
Here I examine the noise spectrum as  another aspect 
of the singular role of Coulomb interaction on collective phenomena
(plasmons, noise) which depends critically  on the dimensionality.

The noise spectrum is quite different in charged and neutral liquids. The equilibrium Johnson-Nyquist
noise (JNN)\cite{Nyquist}
 in an electrical conductor (with a screening length smaller than any size of the conductor)
 is independent of frequency (white noise)  up to the very high frequency
(the smaller of the elastic scattering rate $1/\tau$ and 
the Maxwell relaxation frequency $4\pi\sigma$)\cite{plasma};
while in neutral liquids, the noise becomes frequency-dependent above the ``Thouless'' frequency
$D/L^2$ ($D$ is a diffusion constant and $L$ is the distance between points). The difference 
is due to the screening in charged liquids and depends on the dimensionality of
the conductor. I show here that
for electrical wires having a transverse size ($a$) smaller than the screening 
length $\lambda_D$ (here referred to as ``narrow'' wires), the JNN decays 
as  $\omega^{-\frac{1}{2}}$ above the ``Thouless''
frequency.
Similarly, for two-dimensional films having a thickness smaller
than the screening length the JNN decreases as $1/\omega$ above the characteristic 
frequency $\sigma_{2D}/L$.


To calculate the fluctuations of the electrochemical potential,
 we need to relate it to  the coupled
fluctuations of charge density and currents. We start by writing the continuity equation and
the equation for the current valid in the hydrodynamic limit\cite{Nozieres}:

\beq
\frac{\partial \rho}{\partial t} + div(\vec{j})=0;~~\vec{j}=
\sigma \vec{E}^{tot} - D \vec{\nabla} \rho. \n
\eeq

For self-consistency, we need to account for the potential induced by the fluctuation of charge density:
$\phi_{q,\omega}^{ind}= u_1(q)  \rho_{q,\omega}$ (Coulomb's law). 
If we consider a conductor
with transverse dimensions ($a$) smaller 
than the screening length $\lambda_D$, then $u_1(q)= 2 ln\frac{1}{qa}$ is
a one-dimensional Coulomb potential ($q$ is a wave vector along the one-dimensional conductor).
The total potential driving current
is the sum of the external and induced potentials. Thus the full system of equations is: 

\beq
i\omega \rho_{q,\omega}+iq j_{q,\omega}=0,~~j_{q,\omega}=(iq)\sigma \phi_{q,\omega}^{tot}
+(iq)D\rho_{q,\omega}, \n \\
\phi_{q,\omega}^{tot}=\phi_{q,\omega}^{ext}+\phi_{q,\omega}^{ind},
~~\phi_{q,\omega}^{ind}= 2 ln\frac{1}{qa} \rho_{q,\omega}. \n
\eeq

Finally, after some elementary algebra, we can use the above equations to relate the charge density 
variation to the external potential:

\beq
\rho_{q,\omega}=-\frac{\sigma_1 q^2}{-i\omega+
(D+2 \sigma_1 ln(1/qa))q^2}\phi_{q,\omega}^{ext}, \n 
\eeq 
where $\sigma_1=\sigma a^2$ is a one-dimensional conductivity.
The conductivity $\sigma$ and the diffusion constant $D$ are related by the Einstein formula
$\sigma=D\chi_0$, where
the static charge compressibility $\chi_0$ can be expressed through the Debye screening (or Thomas-Fermi)
length  $\chi_0=1/4\pi\lambda_D^2$.
The density-density response function $\chi_{q,\omega}$ is

\beq
\chi_{q,\omega}\equiv \frac{\rho_{q,\omega}}{\phi_{q,\omega}^{ext}}
 = - \frac{D\chi_0 a^2 q^2}{-i\omega+Dq^2(1 +2a^2\chi_0 ln(1/qa)) }.  \n
\eeq

We can now apply the fluctuation dissipation theorem (FDT) to calculate the density fluctuation
spectrum (assuming classical fluctuations,  $\hbar \omega \ll  kT$)\cite{classical}:

\beq
<\mid\delta \rho_{q,\omega} \mid^2>=\frac{\hbar}{\pi} Im\chi_{q,\omega} 
coth(\frac{\hbar \omega}{2kT}) \cong
\frac{2 kT}{\pi \omega} Im\chi_{q,\omega}.  \n
\eeq

The induced potential fluctuations can be expressed through the charge density fluctuations:

\beq
<\mid \phi_{q,\omega}^{ind} \mid^2>= (u_1(q))^2 <\mid\delta \rho_{q,\omega} \mid^2>, \n \\
<\mid \phi_{q,\omega}^{ind} \mid^2>= 
 \frac{2 kT}{\pi} 
\frac{\sigma_1 q^2(2ln\frac{1}{qa})^2}
{\omega^2 +D^2(1+\frac{a^2}{2\pi\lambda_D^2}ln\frac{1}{qa})^2q^4}.
\label{eq:1d} 
\eeq


We can compare this expression for the spectral density of noise in a 1D wire (Eqn.\ref{eq:1d}) 
with the spectral density of potential fluctuations
in bulk three-dimensional charged and neutral liquids.
In the case of a three-dimensional charged liquid, we need to
use the three-dimensional Coulomb potential 
$\phi_{q,\omega}^{ind}=u_3(q)\rho_{q,\omega} =\frac{4 \pi}{ q^2} \rho_{q,\omega}$.
Following the above simple derivation, we get the expression for voltage fluctuations
(it is sufficient for our purposes to consider only longitudinal fluctuations) in a three-dimensional
conductor\cite{2-der}:

\beq
<\mid \phi_{q,\omega}^{ind(3d)} \mid^2>= 
\frac{2 kT}{\pi q^2}\frac{16\pi^2 \sigma}{\omega^2 +(Dq^2+4 \pi\sigma)^2}. \label{eq:charged}
\eeq

 In the case of a neutral liquid, there is no
long-range induced potential; therefore, we get the standard density-density response function
and potential fluctuations describing diffusion:

\beq
<\mid \phi_{q,\omega}^{(n)} \mid^2>= \frac{<\mid\delta \rho_{q,\omega} \mid^2>}{\chi_0^2}=
\frac{2T}{\pi \chi_0^2} \frac{D |\chi_0| q^2}{\omega^2 +(Dq^2)^2}. \label{eq:neutral}
\eeq

We can now use the spectral densities (Eqns. \ref{eq:1d}-\ref{eq:neutral}) to calculate the
experimentally measured differential noise between the two ends of the sample, averaged
over transverse modes:

\beq
<\mid \phi_{12}(\omega) \mid^2>= 
\sum_{q_x} 4 \frac{ sin^2(q_x a/2)}{q_x^2 a^2}
\sum_{q_y} 4 \frac{ sin^2(q_y a/2)}{q_y^2 a^2} \n \\
\int \frac{dq_z}{2\pi}~ 4sin^2 \left( \frac{q_zL}{2} \right) 
<\mid \phi (q,\omega) \mid^2>. \label{eq:ends} 
\eeq

The Johnson-Nyquist noise in a three-dimensional conductor (\ref{eq:charged}) can be easily calculated,
because the dominant contribution to the sums comes from the transverse zero mode
($q_x=q_y=0$, corresponding to the uniform density of the liquid along the transverse directions):

\beq
<\mid \phi_{12}^{3d}(\omega) \mid^2> = \frac{2kT}{\pi} \frac{L}{\sigma S} 
\frac{1}{1+ (\frac{\omega}{4\pi\sigma})^2 }. \label{eq:3D} 
\eeq

Such noise is readily interpreted as the noise from a conductor 
 having an internal resistance $R=\frac{L}{\sigma S}$ and an
internal capacitance $C=\frac{ S}{4\pi L}$ connected in parallel\cite{Rytov,quasi}:
$ R(\omega)=R/(1+(RC\omega)^2)$. Remarkably, the Johnson-Nyquist noise is white up to the frequency
$4\pi\sigma$, which is independent of the length of the wire. 
It is important to point out
that the noise can depend on frequency through the frequency dependence 
of the conductivity $\sigma(\omega)$.
For the Drude model of conductivity, the characteristic frequency for fall-off of the conductivity
$\sigma(\omega)$ is then the elastic scattering rate $1/\tau$. 

In the case of a ``one-dimensional'' wire $a<\lambda_D$, we can take into account only one
``zero mode''($q_x=q_y=0$), 
since higher harmonics contribute at frequencies of order $D/a^2 > \sigma$.
If we approximate the weak logarithmic dependence in Eqn.(\ref{eq:1d}) by a constant 
$ln\frac{1}{qa} \rightarrow ln\frac{L}{a}$,
we get an expression similar to Eqn.(\ref{eq:neutral}) with the renormalized diffusion
coefficient $D'\equiv D(1+\frac{a^2}{2\pi\lambda_D^2} ln(L/a))$.
 Thus the frequency dependence of noise
for a ``one-dimensional'' wire is the same as for a neutral liquid.
 This result is to be expected, since the screening is not efficient in one dimension.  
The integral over wave vector $q_z$ can be evaluated explicitly assuming for simplicity
$a^2 ln(L/a) \gg 2\pi \lambda_D^2$.
The Johnson-Nyquist noise for a narrow wire is 

\begin{eqnarray}
&&<\mid \phi_{12}^{1d}(\omega) \mid^2> = 2kT  \frac{ L}{\sigma_1 \theta}
(1-e^{-\theta}(cos\theta-sin\theta)), \label{eq:1d-neutral} 
\end{eqnarray}
where 
$\theta=(\omega/2\omega_0)^{1/2}$ and $\omega_0=D'/L^2$ is the natural diffusion frequency.
The expression in the bracket of Eqn. \ref{eq:1d-neutral} is always positive as it must be, 
and the oscillating nature of the second term
is due to the ``resonant'' contributions of the longitudinal ``diffusion modes''.
From the above expression for noise in a one-dimensional wire, we see that it is approximately white
up to the ``Thouless'' frequency $\omega_0$ and equal to $4 kT \frac{L}{\sigma_1}$.
It decays above this frequency as $1/\sqrt{\omega}$.
The same frequency dependence (with $D' \equiv D$) is expected 
for the fluctuations of the chemical potential
between two points in a  narrow vessel ($\omega \ll D/a^2$) of neutral liquid.
In fact, it is the classical result for any quantity (such as temperature, density) obeying
a diffusion process that does not have long-range correlations.\cite{voss} 


The noise in a two-dimensional film ($a<\lambda_D$,  $L_z, L_\perp \gg \lambda_D$)
can be calculated likewise using the above formalism. 
The noise is measured along the $z$ direction, $L_\perp$ and $x$ are the transverse width and 
the transverse coordinate of
the film, respectively, and $a$ is the thickness.
Using the corresponding expressions for 2D Coulomb potential  
$u_2(q)=\frac{2\pi}{ q}$ and the two-dimensional conductivity $\sigma_{2D}=\sigma a$,
the spectral density of 2D noise is

\beq
&&<\mid \phi_{q,\omega}^{ind(2d)} \mid^2>= 
\frac{2kT}{\pi} \frac{4 \pi^2 \sigma_{2D}}{\omega^2 +(Dq^2+2 \pi\sigma_{2D}q)^2}. 
\label{eq:2D-charged}
\eeq

Since the in-plane dimensions of the film are much larger than the screening length,
we can simplify the denominator of the above equation by neglecting the term $Dq^2$
(since $q \ll \frac{\sigma_{2D}}{D}=\frac{a}{4\pi\lambda_D}
\frac{1}{\lambda_D}$, then $Dq^2 \ll \sigma_{2D} q$).
The differential noise between the two ends of the two-dimensional strip, 
averaged over the transverse modes, is

\begin{eqnarray}
<\mid \phi_{12}^{(2d)}(\omega) \mid^2>=
\int \frac{dq_z}{2\pi}~ 4sin^2 \left( \frac{q_zL_z}{2} \right) \n\\ 
\sum_{q_x} 4 \frac{ sin^2(q_x L_\perp/2)}{q_x^2 L_\perp^2} 
<\mid \phi_{q,\omega}^{ind(2d)} \mid^2>.   \label{eq:2D-ends} 
\end{eqnarray}

The integration (if the sum can be approximated by the integral)
 over the transverse dimension $x$ can be done exactly, and
we get the expression:

\begin{eqnarray}
<\mid \phi_{12}^{(2d)}&&(\omega) \mid^2>=
\frac{2kT}{\pi \sigma_{2D}}
\int \frac{dq_z}{2\pi}~ 4sin^2 \left( \frac{q_zL_z}{2} \right) \n\\ 
&&\int \frac{dq_x}{2\pi}~ 4\frac{ sin^2(q_x L_\perp/2)}{q_x^2 L_\perp^2}
\frac{1}{(b^2+q_z^2)+q_x^2}= \n\\
&&=\frac{2kT}{\pi \sigma_{2D}}
\int \frac{dq_z}{2\pi}~ 4sin^2 \left( \frac{q_zL_z}{2} \right) F(q_z). \n
\end{eqnarray}
\beq 
F(q_z)\equiv \frac{1}{2L_\perp^2(q_z^2+b^2)} 
\left(2L_\perp-\frac{1-exp(-2L_\perp\sqrt{q_z^2+b^2})}
{\sqrt{q_z^2+b^2}} \right),\n
\eeq
where $b\equiv\frac{\omega}{2\pi\sigma_{2D}}$ is the inverse of the characteristic length scale of the
problem. The limiting expressions for the function $F(q_z)$ are:

\[ F(q_z)\simeq \left\{ \begin{array}{ll}
                     \frac{1}{L_\perp(q_z^2+b^2)} &\mbox{if $\sqrt{q_z^2+b^2}L_\perp \gg 1$}\\
                     \frac{1}{\sqrt{q_z^2+b^2}}    &\mbox{if $\sqrt{q_z^2+b^2}L_\perp \ll 1$  }
                   \end{array}
           \right. \]

The integral over wave vector $q_z$ can be taken in such limiting cases.
But a careful analysis shows that for the case $\sqrt{q_z^2+b^2}L_\perp \ll 1$  the summation over
the transverse modes $q_x$ cannot be approximated by the integral. The main contribution
actually comes from the ``zero mode'' $q_x=0$ in spite of the condition $L_\perp \gg \lambda_D$.
Taking the above considerations into account,  the answer for the noise
across two-dimensional film is given below.  

\[<\mid \phi_{12}^{(2d)}(\omega) \mid^2> \simeq \frac{2kT}{\pi\sigma_{2D}}* \left\{ \begin{array}{ll}
        \frac{\sigma_{2D}}{\omega L_\perp} &\mbox{if $\omega \gg \frac{2\pi\sigma_{2D}}{L_z},
                                                                 \frac{\sigma_{2D}}{L_\perp} $}\\
        \frac{L_z}{L_\perp}         &\mbox{if $\omega \ll \frac{\sigma_{2D}}{max(L_z,L_\perp)} $  }\\
                               \end{array}
           \right. \]

In some cases, the integral can be expressed through Bessel functions, but only the final asymptotic
expressions are of interest here. As pointed above  
when $L_\perp \ll \frac{\sigma_{2D}}{\omega}$, the main contribution 
(after the integration over $q_z$) calculated from the ``zero mode'' $q_x=0$
is proportional to $L_z/L_\perp$, while the estimate of the integral over the higher harmonics of $q_x$
is smaller and proportional to $ln\frac{L_z}{\lambda_D}$ (if $L_z \gg L_\perp$).
At the frequencies of the order of the 3D Maxwell frequency, the film cannot be considered as
two-dimensional and transverse harmonics other than the ``zero-frequency'' one ($q_y=0$) 
must be taken into account.


It is very interesting that for frequencies above $\frac{\sigma_{2D}}{L_{\perp,z}}=
\sigma\frac{a}{L_{\perp,z}}$ frequency and below the Maxwell relaxation frequency $4\pi\sigma$
the noise is {\it universal}
 (independent of the material specific conductivity $\sigma_{2D}$
and dependent only on the transverse size $L_\perp$) and equal to $\frac{4kT}{\omega L_\perp}$. 

By the FDT the noise is proportional
to the total dissipation which is the product of the dissipation per unit length and the 
characteristic dissipative length scale. The dissipation per unit length is inversely proportional
to the material specific conductivity $\sigma_{2D}$.
At the low frequencies, the dissipative length scale is set by the longitudinal size $L_z$ of the sample,
therefore the noise is proportional to $L_z/\sigma_{2D}$. 
At the high frequencies, as soon as
the length scale $\sigma_{2D}/\omega$ becomes smaller than the longitudinal size $L_z$, the
dissipative length scale is set by this length $\sigma_{2D}/\omega$. 
Therefore the high frequency noise becomes independent of the conductivity.
{\it The universality of the noise} at high frequencies is special to the two-dimensional
situation and is due to dimensional reasons. Only in two dimensions the length scale is 
given by the simple ratio $\sigma_{2D}/\omega$
of the conductivity $\sigma_{2D}$ (or the conductance) and the frequency $\omega$.

At the low frequencies ($\omega \rightarrow 0$),
the noise 
has the standard form consistent with the fluctuation-dissipation theorem\cite{callen} 
applied to the whole sample

\beq
<\mid \phi_{12}^{(2d)} \mid^2> \simeq \frac{2}{\pi} kT R, \label{eq:2D}
\eeq 
where $R=\frac{L_z}{\sigma_{2D}L_\perp}$ is the dc resistance of the film. 
The noise in the 3D wire (Eqn. \ref{eq:3D})
 and the 1D wire (Eqn. \ref{eq:1d-neutral})
is consistent with the FDT as well.


The fluctuation-dissipation theorem  applied to the whole sample relates the voltage noise
between the ends of the sample to the real part of the impedance $Z(\omega)$ of the sample. 
At zero frequency
the capacitive  part of the impedance is always short-circuited by the dissipative part
(the resistance).
The resistance of the wire in all considered cases 
is expressed through the geometrical sizes, as it can be seen from 
the Eqns. (\ref{eq:3D},\ref{eq:1d-neutral},\ref{eq:2D}).
If the resistance is measured
from the zero frequency expression of the noise, then the effective capacitance $C$ of the sample
can be measured  from the high-frequency ($\omega RC \gg 1$) expression of the noise:
$ Re Z(\omega)\simeq\frac{R}{(\omega RC)^2}$.
We can use this equation to interpret the high-frequency noise in the 1D wire
and the 2D strip and to write the expressions for the effective capacitances
of the corresponding wires. 
In case of the 3D wire (with the well screened Coulomb interaction), the sample
has  a constant (frequency independent) capacitance $C=\frac{S}{4\pi L}$. 
The effective capacitance of the 2D strip at the high frequencies 
($\sigma \gg \omega \gg \sigma_{2D}/L_z$) is

\beq
C_{2D}\simeq L_\perp \sqrt{\frac{\sigma_{2D}}{\omega L_z}}=
\frac{L_\perp a}{L_z} \sqrt{\frac{\sigma L_z}{\omega a}}\gg \frac{L_\perp a}{L_z}. 
\eeq
The effective capacitance of the 1D wire at the frequencies ($\sigma \gg \omega  \gg D/L^2$) is

\beq
&&C_{1D}\simeq \left(\frac{\sigma}{\omega} \right)^{3/4} \frac{a^2}{2\pi^{3/4} \lambda_D^{1/2} L^{1/2}}=
\n \\
&&=\frac{1}{2\pi^{3/4}} \left(\frac{L}{\lambda_D} \right)^{1/2} \left(\frac{\sigma}{\omega} \right)^{3/4}
\frac{a^2}{L}  \gg \frac{a^2}{L}.
\eeq
In the both 1D and 2D wires, the effective capacitances are frequency dependent and 
much larger than the standard geometrical
capacitances, because the Coulomb interaction is not completely screened and non-local.


Since the noise has a frequency dependent form, by FDT it implies the same frequency dependence
of the real part of the impedance $Z(\omega)$. The measurement of the complex impedance 
can be more straightforward way to access the predicted frequency dependencies of noise
than a direct measurement of noise.


The nature of the relaxation of a random potential fluctuation is quite different in charged
and neutral liquids. In charged three-dimensional liquids, it is essentially 
the fast process of screening, and
in neutral liquids it is the process of diffusion. 
The appropriate physical picture
of fluctuations in a three-dimensional  electrical conductor is 
that charge fluctuations relax on a very fast
time scale $1/4\pi\sigma$, producing quasi-homogeneous current fluctuations. 
In  one-dimensional systems such as narrow wires,
the Coulomb interaction does not cause long-range correlations; therefore, the noise in a narrow conductor
is similar to the noise spectra in neutral systems. 
The difference
in the chemical potential between two points is relaxed through diffusion on a  characteristic time scale
$L^2/D$
quadratically dependent on the length $L$ between points.
The situation of two-dimensional noise is intermediate, and
the characteristic time scale $L/\sigma_{2D}$ of the relaxation of 
the voltage fluctuation difference between
two points is linearly dependent on the distance $L$. 

The noise spectrum and the spectrum of collective modes  are closely related. Since the
spectrum of collective modes (plasmons) is given by the zeros of 
the dielectric constant $\epsilon(q,\omega)$,
they give rise to the dominant contribution to the noise spectrum 
which is proportional to the  $Im\frac{1}{\epsilon(q,\omega)}$ (see
the comment\cite{2-der}). At the end, both the frequency dependence of the noise and the dispersion
of the collective modes depend essentially only on the effective dimensionality of the Coulomb 
interaction.

The experimental observation of the predicted noise properties is feasible 
in semiconducting materials
having a low carrier concentration.\cite{validity} The screening length 
$\lambda_D$ in such materials
\cite{epi} 
can be as large as $10^{-4} cm$.
In metals, both the elastic rate $1/\tau$ and the Maxwell frequency $4\pi\sigma$ are
high and difficult to observe,
while the typical screening length for a metal 
is of order of $10^{-8}cm$.
In fact, with current experimental techniques 
(see Reference\cite{Naveh} for a review of experiments), even the high Maxwell relaxation frequency
crossover $4\pi\sigma$
can be observed in ``wide'' wires ($a \gg \lambda_D$, the situation almost always encountered)
 with poor conductivity.
By a convenient choice of the mobilities of the semiconductor materials and
their size $L$, the ``Thouless frequency'' $D/L^2$ and the two-dimensional ``relaxation frequency'' 
$\sigma_{2D}/L$ should be accessible. Several other experimental low-dimensional systems
satisfying the condition of the absence of screening can be suggested.

The contacts to the external leads (and associated boundary conditions) are not considered
explicitly in this paper. It is assumed that the main source of noise is the bulk of a wire
or a film, and the contacts have a resistance much lower than a bulk system.

The question of the frequency dependence of equilibrium and ``shot'' noise  
was raised recently by Y. Naveh {\it et al}.\cite{Naveh} Special geometries 
with external screening
were suggested to observe the Maxwell and Thouless crossover frequencies. The above calculation shows
that the crossover at the Maxwell relaxation frequency is a general property of Coulomb systems
 and should be observed independently of geometry and length $L$ for ``wide'' wires. 
Moreover, for ``narrow'' wires ($a < \lambda_D$)
the Thouless frequency crossover should be seen independently of ``external screening'' by
electrodes or the ground plane.

In conclusion, the noise in narrow wires ($a<\lambda_D$) becomes frequency-dependent
starting from the low frequency $D/L^2$ (quite similar to simple diffusion systems), although
in wide conductors, the noise is white up to the smaller of the frequencies $4\pi\sigma$ or $1/\tau$.
In two-dimensional films, the Johnson-Nyquist noise has a universal $1/\omega$ spectrum
in the wide range of frequencies $\sigma \gg \omega \gg \sigma\frac{a}{L}$.

This work was supported by the National Science Foundation through the Science and Technology
Center for Superconductivity (Grant No. DMR-91-20000).
I thank  A. Leggett and M. Weissman for helpful discussions. I am grateful to A. Leggett,
R. Ramazashvili and H. Westfahl
for the useful remarks and the careful reading of the manuscript.

\end{multicols}


\begin{references}

\bibitem{Nyquist}
M.B. Johnson, Phys. Rev. {\bf 29}, 367 (1927), H. Nyquist, Phys. Rev. {\bf 32}, 110 (1928).

\bibitem{plasma}
The Maxwell frequency $4\pi\sigma$ can be expressed through the three-dimensional plasma
frequency $\omega_p$ and the elastic scattering time $\tau$: $4\pi\sigma=\omega_p^2\tau$.
The threshold frequencies $D/L^2$ and $\sigma_{2D}/L$ are much smaller than the three-dimensional
Maxwell frequency $4\pi\sigma$: $4\pi\sigma \gg D/L^2=4\pi\sigma (\lambda_D^2/L^2)$ (1D),
$4\pi\sigma \gg \sigma_{2D}/L =4\pi\sigma (a/(4\pi L))$ (2D).  
These threshold frequency $D/L^2$ and $\sigma_{2D}/L$  can be made arbitrarily 
small by increasing the separation between two points $L$. 


\bibitem{Nozieres}
D. Pines, Ph. Nozi\'{e}res, {\it The theory of quantum liquids} (Benjamin, New York, 1966). For brevity,
we omit the sign $\delta$ of small variations in all expressions.

\bibitem{classical}
Except ultra-low temperatures, the Johnson-Nyquist noise is classical at the frequencies
of interest (e.g. a temperature $kT=0.1 K$ corresponds to a frequency $1.6 GHz$).

\bibitem{2-der}
Equation (\ref{eq:charged}) can be also derived in two other equivalent ways, elucidating
the meaning of charge and current fluctuations in a charged liquid. 
One way is to introduce
a stochastic current source $j^{ac}$ 
into the current equation\cite{kogan}:
$j_{q,\omega}=\sigma(iq) \phi_{q,\omega}
+D(iq)\rho_{q,\omega} + j^{ac} $.
This derivation illustrates the role of quasi-homogeneous current fluctuations effectively
screening charge fluctuations. Yet a third derivation explicitly demonstrates the role
of screening. Since
$ \delta \phi^{tot}_{q,\omega}= \frac{4\pi}{\epsilon(q,\omega)\epsilon_0 q^2} \delta \rho_{q,\omega} $,
 the fluctuations of $\delta \phi^{tot}_{q,\omega}$ due to the FDT are given by the expression
 $<|\delta \phi^{tot}_{q,\omega}|^2>=\frac{\hbar}{\pi} Im\frac{u_3 (q)}{\epsilon(q,\omega)} 
coth\frac{\omega}{2kT} $.
The dielectric constant  $\epsilon(q,\omega)$ can be represented
in the hydrodynamic limit\cite{Nozieres} as:
 $ \epsilon(q,\omega)= 1-4\pi \chi^{(n)}(q,\omega)/q^2 =
1+4 \pi \sigma/(i\omega +D q^2).$ 

\bibitem{Rytov}
S.M. Rytov, Yu.A. Kravtsov, and V.I. Tatarskii, ~~{\it Principles of Statistical Radiophysics}, v.3
(Springer-Verlag, Berlin, 1976), p. 167.

\bibitem{quasi}
All calculations here are done in the assumption of a quasi-stationary condition
( the length of a conductor $L$ is much smaller than the electromagnetic wavelength 
$\lambda=c\frac{2\pi}{\omega}$). This condition implies the absence of the retardation
effects for an electromagnetic field profile inside a conductor (it does not imply the
homogeneity of electric field inside a conductor due to screening and diffusion processes).

\bibitem{kogan}
Sh. Kogan,~~{\it Electronic noise and fluctuations in solids}, (Cambridge University Press, Cambridge, 1996);
B.L. Altshuler, A.G. Aronov, and D.E. Khmelnitsky, J. Phys. C {\bf 15}, 7367-86 (1982).

\bibitem{voss}
R. Voss, J. Clark, Phys. Rev. B {\bf 13}, 556-573 (1976).

\bibitem{callen}
H.B. Callen, T.A. Welton, Phys. Rev. B {\bf 83}, 34-40 (1951).

\bibitem{validity}
It is important to point out that the  condition $a<\lambda_D$ can be satisfied,
while ``weak localization'' and Luttinger-liquid  physics are not relevant
for the problem discussed here. If the electron Fermi-wavelength
$\lambda_F$ is much smaller than the diameter $a$ (many transverse channels),
the Luttinger liquid physics is not relevant. If either 
$\sigma_1=\frac{D}{4\pi\lambda_D^2}\frac{\pi a^2}{L} \gg e^2/\hbar$ 
(or equivalently $L \ll L_{loc}=\frac{\hbar}{e^2} D\frac{a^2}{2 \lambda_D^2}$)
or $L_\phi , L_T \ll a$ (the phase coherence length $L_\phi$, the thermal length $L_T$), than
the ``weak localization'' corrections are small. 


\bibitem{epi}
The large high-frequency dielectric constant $\epi$ of a material is favorable for the experimental
observation, since the screening length $\tilde{\lambda_D}$ becomes larger 
($\tilde{\lambda_D}^2=\epi \lambda_D^2$). The threshold frequencies become smaller 
(e.g. $\tilde{\omega_p}^2 =\omega_p^2/\epi$).

\bibitem{Naveh}
Y. Naveh, D. Averin, and K. Likharev, Phys. Rev. Lett. {\bf 79}, 3482 (1997), 
 Phys. Rev. B {\bf 59}, 2848-60 (1999).

\end{references}
\end{document}